\documentclass[aps,prl,twocolumn,superscriptaddress]{revtex4-2}  

\usepackage{mathrsfs,amsmath,amssymb,amsthm,natbib}
\usepackage{fmtcount}
\usepackage[T1]{fontenc}
\usepackage{graphicx,epsfig,latexsym,overpic,color}
\usepackage{threeparttable}
\usepackage{siunitx}
\usepackage{hyperref}
\usepackage{xcolor}
\usepackage{tikz}    
\usepackage{algorithm} 
\usepackage{algpseudocode} 
\usepackage{dcolumn}
\usepackage{bm}
\usepackage{braket}
\usepackage{orcidlink}

\usetikzlibrary{matrix}
\preprint{\today}

\hypersetup{
    colorlinks=true,
    linkcolor=blue,  
    citecolor=blue,  
    urlcolor=blue    
}

\begin{document}
\title{
Toward  {\it Ab Initio} Quantum  Simulations of Atomic Nuclei Using Noisy Qubits
}
\author{Chongji Jiang\,\orcidlink{0009-0001-9508-4969}}
\affiliation{
State Key Laboratory of Nuclear Physics and Technology, School of Physics,
Peking University, Beijing 100871, China
}
\author{Junchen Pei\,\orcidlink{0000-0002-9286-1304}}\email{peij@pku.edu.cn}
\affiliation{
State Key Laboratory of Nuclear Physics and Technology, School of Physics,
Peking University, Beijing 100871, China
}
\affiliation{
Southern Center for Nuclear-Science Theory (SCNT), Institute of Modern Physics, Chinese Academy of Sciences, Huizhou 516000,  China
}
\author{Rongzhe Hu\,\orcidlink{0009-0002-8797-6622}}
\affiliation{
State Key Laboratory of Nuclear Physics and Technology, School of Physics,
Peking University, Beijing 100871, China
}
\author{Shaoliang~Jin\,\orcidlink{0000-0001-5958-8070}}
\affiliation{
State Key Laboratory of Nuclear Physics and Technology, School of Physics,
Peking University, Beijing 100871, China
}
\author{Haoyu Shang\,\orcidlink{0009-0007-1253-4519}}
\affiliation{
State Key Laboratory of Nuclear Physics and Technology, School of Physics,
Peking University, Beijing 100871, China
}
\author{Siqin Fan\,\orcidlink{0000-0002-7400-9492}}
\affiliation{
State Key Laboratory of Nuclear Physics and Technology, School of Physics,
Peking University, Beijing 100871, China
}
\author{Furong Xu\,\orcidlink{0000-0001-6699-0965}}
\affiliation{
State Key Laboratory of Nuclear Physics and Technology, School of Physics,
Peking University, Beijing 100871, China
}
\affiliation{
Southern Center for Nuclear-Science Theory (SCNT), Institute of Modern Physics, Chinese Academy of Sciences, Huizhou 516000,  China
}

\begin{abstract}
Quantum computers are expected to provide a ultimate solver for quantum many-body systems,
although it is a tremendous challenge to achieve that goal on current noisy quantum devices.
This work illustrated quantum simulations of ab initio no-core shell model calculations of $^3$H with chiral two-nucleon and three-nucleon forces.
The measurement costs are remarkably reduced by using the general commutativity measurement together
with the asymptotic optimization. 
In addition, the noise causes serious contaminations of configurations with undesired particle numbers, and
the accuracies are much improved by applying the particle number projected measurement.  
By tackling the efficiency and noise issues, this work demonstrated a substantial step toward ab initio quantum computing of atomic nuclei. 
\end{abstract}


\maketitle

Quantum computers can in principle simulate quantum systems efficiently  using
resources that scale well with the size of the system.
This capability is particularly attractive for solving nuclear many-body problems.
Atomic nuclei are strongly correlated systems mainly due to the repulsive feature of nuclear forces at short distances.
Ab initio nuclear calculations with realistic nuclear forces are highly anticipated
 to address interdisciplinary fundamental questions~\cite{Ye2025} such as to explore the neutron electric dipole moment,  neutrinoless double-beta decays and extremely dense matter in the cosmos, but for classical computers these calculations suffer exponential explosion of computing costs as the nucleon number grows.

Quantum computing of nuclear systems is still in its infancy~\cite{Hagen2018PRL}, although it promises exponential speedup in ab initio calculations~\cite{QPE1}.
The current state of quantum devices is referred as the Noisy Intermediate-Scale Quantum (NISQ) era~\cite{NISQ2022}.
In this context, most existing NISQ algorithms leverage quantum computing within a hybrid quantum-classical framework.
Taking advantage of shallow-depth  circuits, variational quantum eigensolver (VQE)~\cite{VQE2022} and its variants~\cite{VQE2022,NISQ2022} have become some of the most popular NISQ algorithms and have been applied in a wide range of areas.

In this work, ab initio quantum simulation of atomic nuclei  on noisy qubits, with
two- and three-body nuclear forces, has been advanced substantially.
Previously, quantum computing of configuration-interaction shell model (CISM) has been implemented~\cite{Li62022}.
Presently, we take a step forward by implementing quantum simulations of no-core shell model (NCSM)~\cite{Vary2013PPNP},
which is a full-configuration ab initio method using nucleon-nucleon (\textit{NN}) and three-nucleon (3\textit{N}) forces.
NCSM calculations are among the most challenging computing tasks in nuclear physics since degrees of freedom of all nucleons are active, in particular when 3$N$ forces are included, in contrast to CISM with an inert core.
We employ a chiral $NN+3N$ force~\cite{EM500}, which combines a $\text{N}^3\text{LO}$ $NN$ interaction and a $\text{N}^2\text{LO}$  3$N$ interaction.
Both $NN$ and $3N$ forces are consistently evolved to the same low momentum scale of $\lambda=2.0$ {fm$^{-1}$} under the similarity renormalization group transformation~\cite{srg2007}.
 The wave functions are represented using the unitary coupled cluster (UCC) ansatz and are optimized
by the VQE algorithm. The Jordan-Wigner (JW) transformation and Bravyi-Kitaev (BK) transformation
are used to map the Hamiltonian into qubits~\cite{JW-BK}. Besides, the Gray-code encoding is promising but
more complicated to be implemented~\cite{Singh2025}.


\begin{figure*}[t]
\begin{center}
\includegraphics[width=0.8\textwidth]{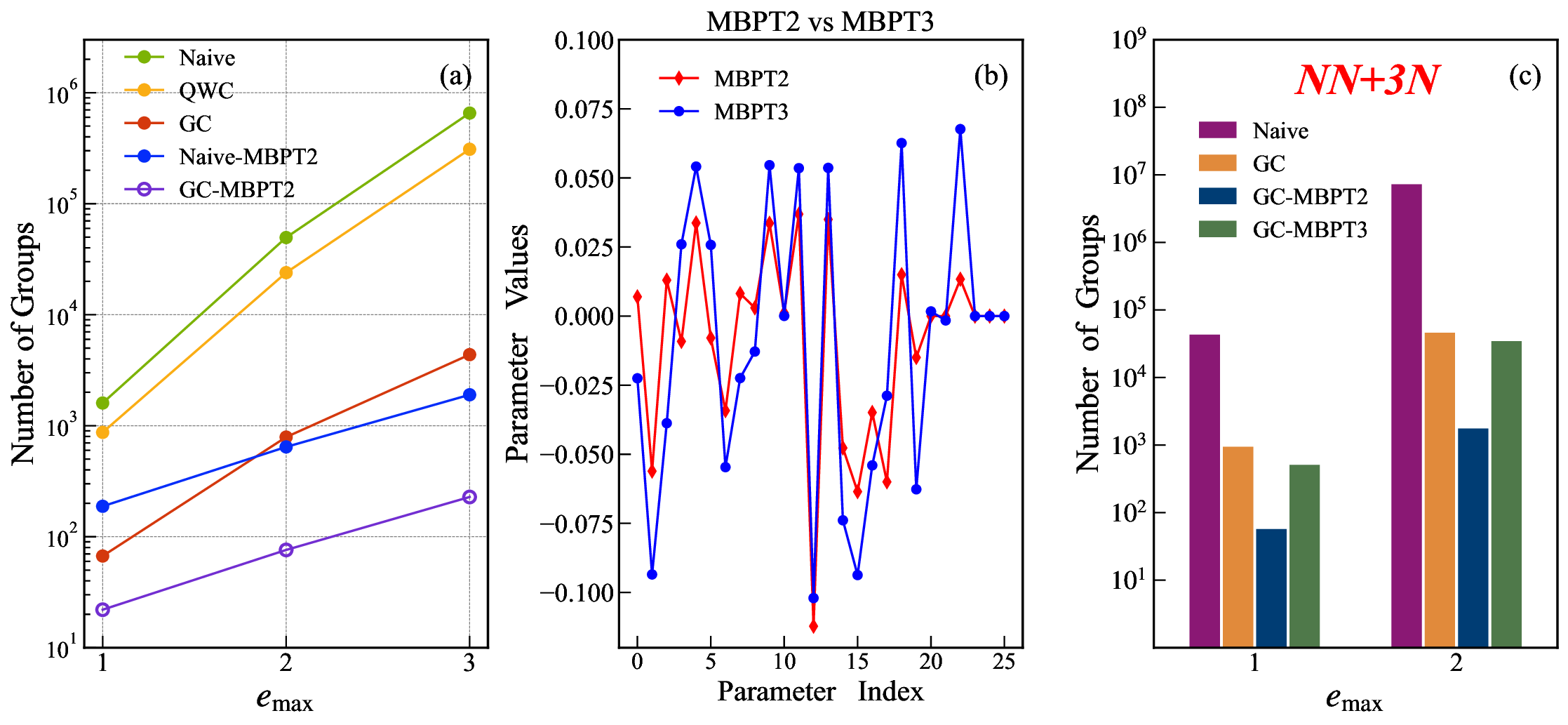}
\caption{
The scalability of measurement costs with increasing model spaces.
(a) Comparison of measurement groups for Hamiltonian with $NN$  forces by different schemes: the naive case without grouping, QWC grouping, GC grouping, MBPT2-related terms with and without GC grouping.
(b) Comparison of parameters obtained by optimizing terms up to MBPT2 and MBPT3 levels, respectively.
(c) Measurement groups for Hamiltonian with $NN$ and 3$N$ forces as the model space increases.
}
\label{FIG1}
\end{center}
\end{figure*}

It is convenient to implement NCSM calculations on qubits using the M-scheme,
in which each qubit corresponds to a single-particle orbit of the Harmonic oscillator (HO) basis.
The number of qubits required increases as a $N_{\rm HO}^3$ polynomial of the number of HO shells.
We take the configuration space of $1s1p$ orbitals for quantum simulations of $^3$H.
Such a model space consists of eight orbitals for protons and neutrons, requiring a total of $16$ qubits.
Within the $1s1p$ model space and UCC up to the triplet excitation, BK mapping requires 2196 single-qubit gates and 2575 two-qubit gates, while 3145 single-qubit gates and 3339 two-qubit gates are required for JW mapping.
The main issue associated with VQE is the measurement of the Hamiltonian, for which a huge number of Pauli strings have to be measured repeatedly, as shown in Fig.~\ref{FIG1}.
The inclusion of 3$N$ force won't make the circuit more complicated, but the measurement cost will be increased tremendously.
In addition to measurement costs, the noise issue becomes serious when a large number of noisy gates are employed for complex problems.
The objective of this work is to tackle
the huge measurement costs and serious noise errors, which are the two major obstacles towards ab initio nuclear quantum computing.

To reduce measurement costs, the simultaneous
grouping measurement method is adopted.
The key of this method is to partition Pauli
strings into groups so that all terms within a group
commute with each other~\cite{Simultaneous2020}.
Fig.~\ref{FIG1}(a) shows that the general commutativity (GC) scheme can reduce the number of grouping partitions by more than one order of magnitude,
compared to the naive case without grouping~\cite{Simultaneous2020}.
The qubitwise commutativity (QWC)~\cite{Simultaneous2020} scheme has been widely used but it is
not so efficient in reducing the grouping numbers.
QWC requires Pauli matrices at each index of two Pauli strings to commute with
each other, while GC only requires even-number of  anticommuting index pairs.
Moreover, it is shown that the decrease of measurement costs by GC becomes more significant as the cutoff of the model space $e_{max}$ increases.

To further reduce the measurement costs, we choose to measure Pauli strings up to
the second-order many-body perturbation (MBPT2) firstly, which is a small subset of  Hamiltonian terms.
The GC groups of MBPT2-related terms are shown in Fig.\ref{FIG1}(a), which are reduced significantly compared to
the number of full terms. Such a reduction effect is more than three order of magnitude at $e_{max}$=3 and becomes
more significant with increasing model spaces.
Note that we are not aiming to calculate binding energies perturbatively but to optimize UCC parameters via MBPT2-related terms.
The estimated variational parameters
at the MBPT2 level are shown in Fig.\ref{FIG1}(b), which are reasonably close to that by optimizing all MBPT3-related terms.
Thus the parameters obtained by optimizing MBPT2-related terms can be seen as a good start for optimizing multi-dimensional exact parameters,
so that the order-by-order optimization can be more efficient.
Indeed, the overlap between MBPT2-guided variational wave functions and exact wave functions is 0.978.

\begin{figure*}[t]
\begin{center}
\includegraphics[width=0.85\textwidth]{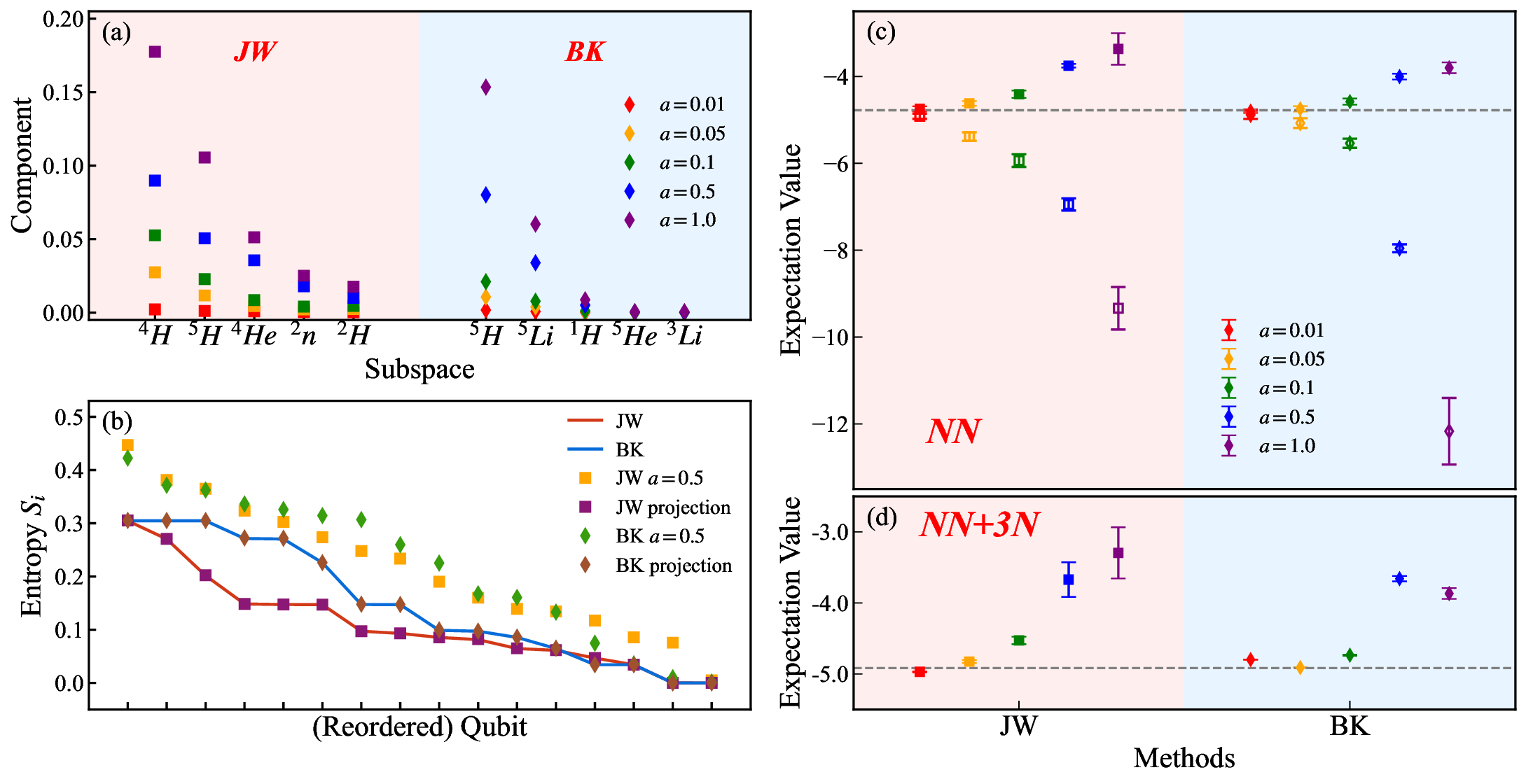}
\caption{
Results with JW and BK mappings as the noise scale $a$ varies.
(a) The magnitude of components with undesired particle numbers due to noise contamination with increasing $a$.
(b) The entanglement entropy of each qubit without noise and with $a$=0.5 are shown.
The entanglement entropy obtained with symmetry projection is also shown.
(c) The total binding energies (in MeV) of $^3$H with only $NN$ forces. The open-symbols indicate the original measured energies.
The solid-symbols indicate the energies with symmetry projection in measurements of diagonal terms, while
other contributions are unchanged.  The exact values from NCSM calculations are shown by the dashed line.
(d) The measured total energies (in MeV) with $NN$ plus 3$N$ forces, in which diagonal terms
are measured with symmetry projection.
}
\label{FIG2}
\end{center}
\end{figure*}

The inclusion of full 3$N$ forces leads to a rapid increase of the number of Pauli strings.
Fig.\ref{FIG1}(c) shows that the number of Pauli strings is larger than that of $NN$ forces in Fig.\ref{FIG1}(a)
by two orders of magnitude at $e_{max}$=2.
Again, we see that the GC scheme can reduce the measurement groups remarkably.
For $e_{max}$=2, the GC scheme can reduce the groups by more than two orders of magnitude.
The GC grouping at the MBPT2 level can reduce the number of groups by
one more order of magnitude at $e_{max}$=2.
In our case, the grouping at the MBPT3 level is about 75\% of all terms, which is
sufficient for UCC ansatz up to triplet excitations.
We expect that the GC grouping scheme and the order-by-order optimization
 can provide a promising solution to the measurement bottleneck of VQE.

In addition to measurement costs,
 it is critical to understand the behaviors of noises
and to mitigate noise errors in the NISQ era. There have been several successful schemes for
error mitigation, such as zero-noise extrapolation (ZNE)~\cite{ZNEprx,Thermal2023}, readout error mitigation~\cite{Thermal2023},
symmetry preservation~\cite{Thermal2023} and verification~\cite{SV2018}, error-correction circuit~\cite{EC2023}, etc.
For ab initio many-body calculations, it is desirable to understand
 how the noises distort many-body correlations and then develop specific remedies
 to improve the accuracy.

To systematically evaluate the performance of the noisy quantum devices under varying noise levels, we introduce a scaling coefficient $a$, where $0<a\leq1$, which uniformly scales all gate error probabilities within the noise model built in Qiskit.
The noise model is derived from real noise data obtained from the IBM quantum device Kyoto~\cite{qiskit2024}, including various noise channels.
This approach allows us to construct a tunable noisy simulator, enabling controlled investigations of the impact of noise on quantum state preparation.
We chose 10,000 shots for each measurement to ensure statistically reliable results.

Quantum computing is expected to be performed within a restricted model space belonging to the full space of $2^N$ states.
 However, the practical model space would be contaminated by the full space gradually due to noises~\cite{NISQ2022}.
 Fig.\ref{FIG2}(a) shows that with increasing noise scales, other undesired components with different particle numbers appear.
 For the JW mapping, the distribution of different particle numbers is broad.
 For the BK mapping, the main redundant components are $^5$H and $^5$Li, with additional two particles.
 The non-conservation of particle numbers would cause serious accuracy problems, in particular, for light nuclei.

To further illustrate the decoherence between noisy qubits, the entanglement entropy among qubits has been calculated,
as shown in Fig.\ref{FIG2}(b). It can be seen that at zero noise scale, the BK mapping results in a more entangled circuit
than JW.  With increasing noise scale, the entanglement entropy is much increased and the entropy of two mappings is close
at the noise scale of 0.5.
Here the additional entropy is caused
by the mixture of other configurations due to noises~\cite{NISQ2022}, but doesn't mean increased entanglement magnitude.
We implement the projection on particle numbers, parity and $z$-component of the angular momentum ($J_z$) on the density matrix to obtain purified $^3$H wave functions.
Then we see that the entropy of each qubit is restored to the zero noise scale.
This can be understood that the entropy is very much related to the one-body reduced density matrix and the orbital occupation probability~\cite{NuclearEntangle2021}.

The final measured results of binding energies of $^3$H with varying noise scales are shown in Fig.\ref{FIG2}(c,d).
The original results using only $NN$ forces are shown as open-symbols in the upper panel.
We see the total energies decreasing seriously with increasing noise scales.
The BK mapping is better at low noise scales but worse at high noise scales,  compared to the JW mapping.
At low noise scales, the BK mapping has a more entangled circuit with fewer gates and results are more accurate.
At high noise scales, the BK mapping with contaminations of $^5$H and $^5$Li results in overestimated binding energies.
To overcome the dominant accuracy issue, the energies related to diagonal Hamiltonian elements are measured with the symmetry projections mentioned above.
This can be achieved efficiently since these terms contain only particle number operators.
Then the resulted total energies are shown as solid-symbols.
We see that the accuracy is much improved. For the BK mapping, the deviation has changed from the original -7.39 MeV at $a$=1.0 to 0.98 MeV.
The actual deviation in the contribution of diagonal elements is decreased from -8.33 MeV to 0.04 MeV.
The diagonal-part energies can be restored by applying symmetry constraints, particularly the conservation of particle numbers.
Similarly, the total energies with 3$N$ forces  are shown in the lower panel.
The deviations due to noises are slightly larger than that with $NN$ forces.
This is because simulations with 3$N$ forces correspond to much more Pauli strings, while
the circuit is the same, compared to the case with only $NN$ forces.
For $^3$H, the energy difference due to full 3$N$ forces is small.
By employing symmetry projections in measurements, we see that the BK mapping is more accurate than JW.


To perform ab initio full-configuration calculations of heavier nuclei on current noisy  quantum computers is still a tremendous challenge.
The main issue is the huge measurement cost, in particular when 3$N$  forces are included, which is the efficiency bottleneck of the VQE algorithm.
We demonstrated that this bottleneck can be much alleviated by using the GC simultaneous measurement. Actually, GC together
with the MBPT2-guided optimization can reduce the measurement cost by three orders of magnitude for $^3$H with 3$N$ forces.
The measurement costs could be even reduced considering the sparsity of the Hamiltonian with increasing model spaces.
Consequently it is practical to perform ab initio quantum computing of light $p$-shell nuclei with the cutoff at $e_{max}$=2 on near-term devices.
In addition to measurement costs, the mitigation of noise errors is crucial, using noisy qubits for complex problems.
For ab initio calculations,  a prominent problem is the contamination of configurations with undesired particle numbers.
This is particularly serious for the contribution of diagonal Hamiltonian elements, and we demonstrated that it can be cured by applying symmetry projection in the measurement.
Then the accuracies of total Hamiltonian energies can be improved by several times. The accuracy could be further improved
by the combined usage of other error mitigation methods.
In conclusion, this work tackled the two most-concerned obstacles,  by remarkably reducing measurement costs and suppressing noise contaminations,  taking a substantial step
towards ab initio quantum computing of light nuclei.
Note that quantum annealing platforms can provide an alternative solution of nuclear systems in addition to VQE~\cite{Annealing}.
It is obvious that more efficient and  noise-resilient algorithms in this direction are still highly anticipated.
\\


\noindent{\bf{Acknowledgments}}\\
This work was supported by  the
 National Key R$\&$D Program of China (Grant No.2023YFA1606403, 2023YFE0101500),
  the National Natural Science Foundation of China under Grants No.12475118, 12335007.
\\

\noindent{\bf{Author Contributions}}\\
C.J. wrote the code and performed theoretical simulations;  J.P. conceived
the project and wrote the manuscript; R.H., S.J., H.S., S.F., and F.X. helped
 ab initio nuclear calculations and  participated in discussions.



\end{document}